\documentclass[a4paper,nobibnotes,nofootinbib]{revtex4}

\usepackage{amsmath,amssymb}
\usepackage{epsfig}
\usepackage{graphicx}
\usepackage{xspace}

\newcommand{\bL}{\begin{Large}}
\newcommand{\eL}{\end{Large}}
\newcommand{\ba}{\begin{eqnarray}}
\newcommand{\ea}{\end{eqnarray}}
\newcommand{\bc}{\begin{center}}
\newcommand{\ec}{\end{center}}
\newcommand{\bfig}{\begin{figure}}
\newcommand{\efig}{\end{figure}}

\newcommand{\g}{\gamma}

\newcommand{\ka}{\kappa}

\newcommand{\la}{\label}

\newcommand{\alp}{\alpha_{\cal P}}

\newcommand{\eps}{\epsilon}

\newcommand{\cN}{{\cal N}}

\newcommand{\rr}[4]{#1, {\it #2 \/}{\bf #3} #4}

\newcommand{\tm}{\ensuremath{\mathrm{|t|}}\xspace}

\begin{document}

\title {Hybrid Pomeron Model of exclusive central diffractive production}

\author{Robi Peschanski}
\email{robi.peschanski@cea.fr}
\affiliation{IPhT, Institut de physique th{\'e}orique, CEA/Saclay,
  91191 Gif-sur-Yvette cedex, France\footnote{%
URA 2306, unit\'e de recherche associ\'ee au CNRS.}}

\author{M. Rangel}\email{rangel@cbpf.br}
\affiliation{LAFEX, Centro Brasileiro de Pesquisas Fis{\'i}cas, Rio de Janeiro, 
Brazil}

\author{C. Royon}\email{royon@hep.saclay.cea.fr}
\affiliation{CEA/IRFU/Service de physique des particules, CEA/Saclay,
  91191 Gif-sur-Yvette cedex, France}
  
\begin{abstract}
Central diffractive production of heavy states (massive dijets, Higgs boson) is 
studied in the exclusive mode using a new {\it Hybrid Pomeron Model} (HPM). Built 
from  Hybrid Pomerons defined by the combination of one hard and one soft color 
exchanges, the model describes well the centrally produced diffractive dijet 
data at the Tevatron. Predictions for the Higgs boson and dijet exclusive 
production 
at the LHC are presented.
\end{abstract}
\maketitle

%%% ----------------------------------------------------------------------

\section{{Introduction: The Hybrid Pomeron}}
Central diffractive production of heavy objects in its $exclusive$ mode (no 
other particle produced in the central rapidity region) appears as a promising 
complementary tool for the study of new particles at the LHC, such as the Higgs 
boson. Indeed, for instance, the mass determination 
can be made quite precise, if both incident protons are detected and measured in 
forward detectors located at 220 and 420 m from the 
interaction point at the LHC~\cite{forward,al00}. One expects to take advantage 
of the absence of 
other  
particles than the decay products in the central rapidity region and some other 
interesting aspects such as the depletion of b-quark production due to the 
helicity rule specific of this production mode~\cite{helicity}. The key problem 
of central 
diffractive production in the exclusive mode is to determine its rate as a 
function of $e.g.$ the Higgs boson mass or the minimum $p_T$ of the jets. 
Experimental results on massive dijet production at the Tevatron has shown 
indirect evidence for exclusive production, by comparison with models of 
$inclusive$ diffractive production. Inclusive models \cite{cdf, Kepka:2007nr} 
agree to 
point out an excess of events over the inclusive spectrum in the kinematical 
region 
where exclusive production is expected to contribute.

The experimental interest of  central exclusive production, in the first place 
for 
the Higgs boson and $\gamma$ induced processes, and the preparation of concrete 
proposals at the LHC is a major 
incentive for theorists to work out  reliable predictions for the production 
cross section, which could serve as a basis for the necessary data simulations. 
This task is not easy since central diffractive processes imply both hard 
subprocesses, related to the high mass of the centrally produced 
states, 
and soft ones which are typical of diffractive events which leave intact the 
initial 
particles -$e.g.$ protons at the LHC. In some sense one could say that central 
diffractive production is expected to combine the ``hardest'' events such as the 
production of massive Higgs bosons or of any high mass object (dijet, 
diphoton...), with the 
``softest'' ones, since the initial particles remain totally intact (up to a 
loss of 
energy not bigger than 10 \%). This reveals the potentially $hybrid$ character 
of 
central diffractive production.

On the theoretical ground, different mechanisms of exclusive central diffraction 
have been proposed since years~\cite{others}, but we will restrict to two 
classes  of models 
which are based on the exchange of colorless objects, in order to take into 
account 
the diffractive property. Indeed, any colored object would generate particle 
production in the whole rapidity interval\footnote{One notable exception is 
the 
Soft Color Interaction (SCI) model \cite{sci} where a colorful exchange is 
compensated by a phenomenological soft color interaction at long distance, which 
generates a gap in rapidity. We do not consider this model in the further 
discussion since it would need modifications to describe the CDF measurement
of the dijet mass fraction~\cite{Kepka:2007nr}.}. One class is  based on the 
exchange 
of two Pomerons, where the Pomeron is the colorless exchange which appear in 
$e.g.$ elastic 
reactions; it can be called the Non Perturbative Model (NPM) and was based on 
a typical soft interaction hypothesis, which comes from the Bialas-Landshoff 
mechanism~\cite{bialas1} originally proposed for central diffusive production. 
It 
has an $inclusive$ version which describes the inclusive diffractive dijet 
production \cite{us}, while its $exclusive$ version has been studied in 
Ref.\cite{us1}. One another class of models is based on the exchange of two 
gluons 
at each vertex for the exclusive production~\cite{khoze} called KMR (from
the author names) in the following. For both models there 
exists a detailed phenomenological discussion (see $e.g. $\cite{Kepka:2007nr}) 
based 
on dedicated simulations.

Let us recall the present status of this physically meaning discussion. The 
inclusive production 
mechanism based on the NPM~\cite{us} gives satisfactory results when compared to 
Tevatron  data.
%\footnote{There is now agreement that the model with a factorizable 
%Pomeron hypothesis \cite{Kepka:2007nr} 
%has been favored w.r.t. the non factorizable 
%one which was initially considered in~\cite{us}.}. 
Using this agreement, the 
extraction of the exclusive component in the DPE framework becomes possible, 
since 
it appears to be necessary to include it in a well-defined region of 
phase-space. When 
comparing~\cite{cdf,Kepka:2007nr} the extracted dijet cross section and spectra 
with the models, it appears 
that the KMR model~\cite{khoze} gives a better description of the results than
NPM~\cite{us}. The main reason is that it takes into account 
the Sudakov suppression factors   
preventing cross section s to include the gluon radiation normally associated 
with 
the production of a massive object. The soft Pomeron exchanges of NPM~\cite{us} 
do not 
contain these perturbative QCD factors and give a too flat distribution as a 
function of the minimum transverse momentum $p_T^{min}$ of the jet 
\cite{Kepka:2007nr}. As a consequence, the prediction for the Higgs boson 
cross section, which was similar for both models for a light Higgs boson 
\cite{Forshaw:2005qp}, has a different form as a function of the Higgs boson 
mass, 
being steeper for the KMR model~\cite{khoze} than for NPM~\cite{us}. It is 
expected that
the NPM model in Ref.~\cite{us} works at low masses (for instance for $\chi_C$ 
production~\cite{chic})
whereas a model including a hard contribution may be valid at higher masses.

Our motivation is to keep the Double Pomeron Exchange (DPE) hypothesis, while 
taking into account the fact 
that the diffractive production process is expected to be a mixture of soft and 
hard 
color exchanges. Indeed, the notion of a hard Pomeron (associated in  QCD with 
the 
summation of ladder diagrams in the leading or next-to-leading logarithmic 
approximation  (LLA) of the perturbative expansion) is a  theoretical result of 
QCD 
\cite{BFKL}.  Moreover it has been successfully compared with data in (hard) 
inclusive diffraction (see, $e.g.$ \cite{Bialas:1997vt,Marquet:2007nf}) and 
exclusive vector-meson 
production \cite{marquet}. The example of heavy vector meson production, in 
particular, is well suited for our approach  since it corresponds to the 
(quasi-) 
elastic production of a heavy state, which can be formulated in the framework of 
a 
hard Pomeron exchange. 

In the theoretical calculations, the hard Pomeron appears to correspond to 
ladder 
diagrams connecting  two exchanged reggeized colored gluons \cite{BFKL}.  
However, 
in central diffractive production, one could expect to have two different 
colored 
exchanges,  one  hard and one soft. It would correspond physically to two time 
scales, one short corresponding to the heavy state production, and one long 
corresponding to the necessary color neutralization.
%\footnote{It would thus be 
%similar, however formulated differently since using the Pomeron formalism, to 
%the 
%SCI mechanism. {\bf I do not understand: SCI is more a model related
%to hadronisation?}}.  
Hence, the qualitative picture of central diffractive production 
which we  formulate is a  DPE process in which each Pomeron exchange at the 
vertex 
would correspond to $hybrid$ Pomerons with two different types of color 
exchanges 
one  soft and one hard. It would correspond to an $effective$ way of summing 
ladder 
diagrams between hard and soft colored reggeized gluons, which precise 
calculation 
remains beyond our scope (and beyond the present knowledge of non perturbative 
QCD 
physics).

The plan of the paper is the following. In the next section, we shall formulate 
the 
Hybrid Pomeron Model (HPM) and determine its parameters  obtained from known 
soft and hard 
Pomeron processes. In section III, we will show its good description of  
exclusive 
dijet production extracted from data at the Tevatron and the prediction for 
Higgs 
Boson and dijet production at the LHC. The last section is for discussions, 
conclusions and an outlook.

%\eject
\section{Formulation: The Hybrid Pomeron Model}

The model, adopting as a starting point the idea of the original 
Bialas-Landshoff formulation, consists in defining effective propagators and 
couplings for the colored exchanges associated with central DPE processes, see  
Fig.\ref{1}. However, by contrast with the original NPM model of 
Ref.\cite{bialas1}, we introduce two 
types of propagators and couplings, depending of its soft or hard character. The 
soft propagator $D_S$ and coupling $G_S$ are exactly those which appear in the 
original description of the Bialas-Landshoff model \cite{bialas1}, themselves 
connected to the soft Pomeron Landshoff-Nachtmann formulation \cite{LN} of the 
elastic cross section, see Fig.\ref{2}. They are  constrained to describe the 
elastic hadronic cross section, which fixes  its parameters.

\begin{figure} [htb]
\begin{center}
\epsfig{file=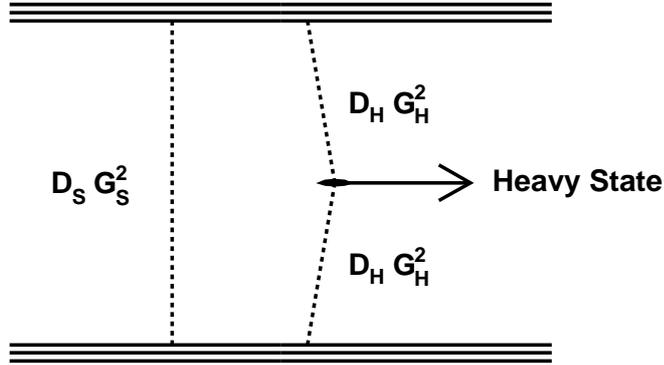, height=0.30\textwidth}
\end{center}
\caption{{\it Hybrid Pomeron Model.} The dotted lines schematically represent 
the 
colored exchanges. They are formulated in terms of effective propagators  for 
soft 
and hard exchanges (resp. $D_S$ and $D_H$) and couplings (resp. $G_S$ and 
$G_H$), 
see text.}
\la{1}
\end{figure}

The new aspect, w.r.t. the original formulation \cite{bialas1}, is to introduce 
similarly effective propagators and couplings  $D_H$ and coupling $G_H$ for hard 
Pomeron processes. Since we are formulating a {\it hybrid Pomeron Model} (HPM), we 
would 
optimally need the resummation of QCD ladder diagrams corresponding to  both 
soft 
and hard colored exchanges. As we see in Fig.~\ref{1}, the hard exchange 
produces
the heavy state object ($D_H G_H^2$) while the colorless aspect of the exchange 
is ensured 
via the emission of a soft additional gluon ($D_S G_S^2$). This means that most 
of
the available momentum is carried away by one of the gluon, the hard one, while 
the soft one
carries only a very small fraction of the proton momentum. 
%We already understand from which data the 
%parameters of the hybrid model will come: 
The hard part of the HPM model will be based on hard
physics measured at HERA (for instance the proton structure function
$F_2$) while the soft part will be based on usual soft cross section
measurements. 
In the case of simple elementary gluon exchanges,  as 
developed in the  model \cite{khoze}, the problem is perturbatively 
tractable\footnote{At least partly, since the considered  models have to correct 
for 
the rapidity gap survival probability, corresponding to the interaction between 
incident 
particles~\cite{sp,kupco}.}, since the loop kinematics enforces a (semi-) 
perturbative calculation. 
However, when considering Pomeron ladders, the gluon loop constraint, 
characteristic 
of the mechanism of \cite{khoze} does not hold and thus one relies on the 
Bialas-Landshoff modified picture in order to include both soft and hard 
effective color exchanges, see 
Fig.\ref{1}. Hence our proposal is to start  with  the description of hard 
Pomeron 
scattering in terms of effective hard colored propagators $D_H$ and couplings 
$G_H$, 
in the same way as for the soft color exchanges in \cite{bialas1}.

For this sake we consider the well-known dipole-proton amplitudes which appears 
in 
the QCD description of many hard processes. They will be  used to determine the 
effective propagators and couplings. In that sense, it is possible to fix the 
parameters
of the model using hard physics measurements at HERA, especially from the
measurements of the proton structure function and the vector meson production
cross sections.
In this basic process, a 
dipole of size $r$ experiences an elastic scattering with the proton. Since this 
dipole-proton amplitude, corresponding to an hard Pomeron exchange, appears in 
the 
formulation of different observables, its parameters are well determined, and 
thus 
gives the possibility to define the appropriate  hard propagators $D_H$ and 
couplings 
$G_H$, in  the same way as was done for the soft ones, but with the advantage 
that 
we have a theoretical control on its precise QCD formulation.

A comment has to be made at this stage. The main new aspect of HPM is to 
introduce a formulation for hard color exchanges. Since it is a 
phenomenological  effective description of diagrams going beyond elementary 
gluon 
exchanges, it aims at  keeping the physical image of two different time scales 
and 
thus of two different types of effective propagators. Hence the virtuality 
associated with the hard color exchanges cannot be transferred to the other 
color 
exchange through the loop kinematics, as is the case in the model \cite{khoze}. 
On 
the other hand the inclusion of hard color exchanges in the DPE formulation is 
expected to  (and indeed will, as we shall see) correct the drawbacks of the 
initial 
soft model.

\subsection{\textbf{Soft color exchange}}

\begin{figure} [t]
\begin{center}
\epsfig{file=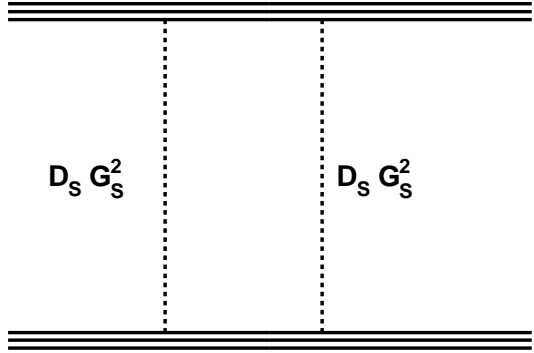, height=0.22\textheight}
\end{center}
\caption{{\it Proton-proton elastic scattering in the Landshoff-Nachtmann 
formulation.} The elastic amplitude is described by two-color exchanges 
associated 
with ``non-perturbative'' gluon propagators $D_S$ and couplings $G_S$ 
\cite{bialas1}.}
\label{2}
\end{figure}

We evaluate the non-perturbative gluon propagator from the elastic
proton-proton data, see Fig.~\ref{2}. Following Landshoff-Nachtmann 
proposal~\cite{LN}, the
elastic hadron-hadron scattering is  represented by
the contributions of elastic valence  quark scattering mediated by a 
non-perturbative model for gluon exchange. The  elastic quark-quark amplitude in 
terms of soft propagator and coupling writes\footnote{We have incorporated the 
Regge 
factors due to reggeization \cite{bialas1} in the definition of the 
propagators.}
\begin{equation}
A_{qq} \equiv G_S^{2}\ D_S = s^{\alp (t)}\ G_s^{2}\ D_S^{(0)}\  e^{\ \frac 
t{\mu_S^{2}}}\ ,
\la{ampli}
\end{equation}
where $s$ is the total c.o.m. energy, $t$  the transfer quadrimoment squared 
whose 
dependence is approximated by an exponential slope given by $\mu_S$ and 
\begin{equation}
\alp (t) \equiv \alp(0) + \alp' \log s = 1+\eps + \alp' \log s
\la{regge}
\end{equation}
is the soft Pomeron Regge trajectory \cite{pom}, with
$\eps \sim .08$ being  the  Pomeron ``anomalous intercept''. Note that we have 
incorporated the factors due to reggeization \cite{bialas1} in the definition of 
the propagators. This is required in order to take into account the different 
Regge 
parameters (and in particular the known different energy dependence) between the 
soft and hard Pomeron ingredients. In other terms the {\it hybrid Pomeron} will have 
an 
intermediate energy dependence compared to the soft and the hard Pomeron's ones.

%\vspace{1cm}

All in all, the differential elastic hadronic cross section, from which the 
relevant 
parameters  will be obtained, is given in a suitable normalization, by 

\begin{equation}
\frac{d\sigma}{dt} \equiv \frac{1}{4\pi s^2}\left\{9  A_{qq}\right\}^2 =  
\left| 3\beta \right|^{4} s^{2\alp(0)-2}
\exp \left[( 4 b + 2\alp \log s )\ t  \right]\ ,
\la{sigma}
\end{equation}
where the parameters $\beta$ can be obtained \cite{bialas1} from the total 
cross section  and  $b$ from the elastic form factor or equivalently, from the 
differential cross section s. Note that
the factor $\left\{9  A_{qq}\right\}$ comes from the number of valence quark 
combinations, which are 
considered independent in the  Landshoff-Nachtmann formulation.

Using the effective propagator and coupling formulation of \eqref{ampli}, one 
finally determines

\begin{equation}
\begin{split}
&\mu_{S}^{-2} = 2b + \alpha'\log s\\
&\left[G_{S}^{2}\ D_{S}^{(0)}\right]^2 = 8 \beta^2 s^{2\epsilon} \exp (4b + 
2\alpha'\log s)
\ .
\end{split}
\end{equation}

\eject

%%% ----------------------------------------------------------------------
\subsection{\textbf{Hard color exchange}}

For the definition of the propagator and coupling of the hard color exchange, we 
will use the well-known hard Pomeron for the dipole proton elastic amplitude 
calculated from perturbative QCD using as a starting point the Balitsky Fadin 
Kuraev 
Lipatov (BFKL) equation~\cite{BFKL}. It is convenient to open the possibility of 
saturation effects, even if they are not expected to be important in the 
kinematical domain we are interested in. Indeed, this form of the dipole proton 
amplitude (eventually modified by 
saturation contributions) has been proven to be phenomenologically successful in 
the 
description of proton total and diffractive structure functions 
\cite{Bialas:1997vt} and, more importantly for our analysis, 
for structure function $F_2$ measured at HERA including the charm 
contribution~\cite{soyez}  for vector meson elastic 
differential cross section ~\cite{marquet} and for inelastic diffraction 
\cite{Marquet:2007nf}, which will be used for  parameter 
fixing. Hence the model we will adopt for dipole-proton elastic
scattering contains saturation effects and \tm dependence~\cite{marquet}.

\begin{figure} [hbt]
\begin{center}
\epsfig{file=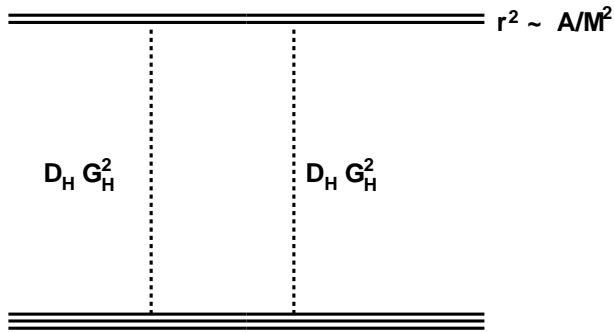, height=0.2\textheight}
\end{center}
\caption{{\it Dipole-proton elastic amplitude.} The elastic amplitude is 
described 
by two-color exchanges associated with hard colored propagators $D_H$ and 
couplings 
$G_H$, see text. As well-known, the dipole size is approximately related to the  
vector meson mass $r^2 \propto M^{-2}.$}
\la{3}
\end{figure}

We start with the following amplitude in terms of the BFKL kernel. One writes
\begin{equation} 
\cN (r,Y)=\int_{\cal C}
\frac{d\gamma}{2i\pi}\, \cN_0(\gamma)\,r^{2\g}
\exp \left\{ {\bar\alpha\chi(\gamma)} Y\right\}\ ,
\la{equa1}
\end{equation}
where $\chi(\gamma)$ is the Mellin transform of the BFKL
kernel \cite{BFKL}. $\cN_0(\gamma)$ contains information on the coupling to the 
proton and other normalization contributions. 

The effect of saturation, through nonlinear damping factors,  is known 
\cite{munier} 
to select a $critical$ value $\g_c.$ this corresponds to a ``anomalous 
dimension'' 
$d_c = \g_c\!-\!1$ which is characteristic of a (saturation-corrected) hard 
Pomeron.
In a more concrete way, the authors of \cite{marquet} make use of a model for 
the 
dipole-proton amplitude \cite{Iancu:2003ge} which successfully describes the
precise proton structure function data.
\begin{eqnarray}
\cN (r,Y) & = &  N_{0}\ (P_{H})^{\g_c} 
\exp\left( - \frac{\log^2(P_H)}{2\kappa \lambda Y} \right)
\exp(-B|t|) \nonumber\\
P_H & = & (r /r_{S})^{2} \nonumber \\
&  & \nonumber \\
r_{S}^{2} & = & r_{0}^{2} \exp(-\lambda Y)\ ,
\la{equa2}\end{eqnarray}
where $Q_S \equiv 2/r_s$ is the well-known saturation scale, the term 
$\exp(-B|t|)$  
has been added in order to take into account the  momentum transfer dependence 
in  
vector meson  production \cite{marquet}.  The exponential term in \eqref{equa2} 
takes into account the contribution from the kernel variation  around the  
saddle-point. Note that this amplitude works \cite{Iancu:2003ge}
in the region $P_{H} < 1$, which is safely true in our case. The saturation
corrections are expected to be negligible in that region.

It is  important to notice at this stage that we need to consider   {\it 
amputated} 
amplitudes, that is multiplying the expressions (\ref{equa1},\ref{equa2}) of the 
dipole-proton  amplitude by  a factor $r^{-2}.$ Indeed,  we have to remove from 
the 
usual dipole proton amplitude the factor corresponding to the   geometrical 
dimension of the dipole cross section  proportional to $r^{2}$,
or on other words the gluon dipole coupling. 
This factor 
has to be removed in order to define properly the couplings and propagators of 
the 
hard effective color exchanges, which would be valid for any massive state. We 
are 
interested in applying our formalism to the exclusive production of  massive 
dijets 
or the Higgs boson and thus have to switch from a dipole state to the  wave 
function 
coresponding to the heavy state under study.

In the kinematical configuration of central diffractive production, we have at 
each 
hard color exchange vertex (see Fig.~\ref{1})
\begin{eqnarray}
Y = -log(\xi) \ ,{\rm \ and\ } r^2 \sim \frac{A}{M^2}
\la{Y}\end{eqnarray}
where we used a very simple\footnote{More refined wave function analyses are 
straightforward extensions of our formalism.} relation (with $A \sim 7$ 
phenomenologically) between the mass $M$ of the heavy state and the 
corresponding 
dipole size $r$ to be considered.
The normalization factors  $N_{0}$ for the amplitude and the scale $r_{0}$  are 
also determined  phenomenologically from HERA data.
For dipole-proton scattering, we assume also a dominance of ``valence'' 
quark-quark 
scattering
and 6 quark-quark combinations
are allowed. 

The main characteristic feature of the hard Pomeron by contrast with the soft 
one is 
that it  has a non trivial dependence
on  $Y$ and $r$ (translating into a non trivial $\xi$ and $M$ dependence in the 
central diffraction kinematics) through the anomalous dimension $\g_c$. Indeed, 
this perturbative QCD dependence plays the role of the Sudakov form factors in a 
BFKL-like model. It acquires also a 
different, faster, energy dependence through the dependence on $P_H$ in 
\eqref{equa2}. 

Using concretely the parameters from the fit to the HERA 
data~\cite{soyez,marquet}, 
one finds the following 
expression for the couplings and propagators of the hard gluon exchanges:
%\begin{center}
\begin{eqnarray} \label{eq:shgluon}
\mu_H &=&  0.5 \nonumber\\
\left[G_{H}^{2} D^{(0)}_{H}\right]^{2}  &=& 
\frac{8\pi}{6\mu_H^2} \ \times \ 2\pi R_p^2 N_0 \times {r}^{-2}  
\times (P_{H})^{\g_C} 
\exp\left( \frac{\log^{2}(P_{H})}{2\ka\lambda \log(\xi)} \right)\ ,
\end{eqnarray}
%\end{center}
where we have used the values for $\ka,\lambda,B$ (see formulae~\eqref{equa2})  
taken from the 
phenomenological analysis~\cite{soyez,marquet} of massive vector mesons, charm
and stucture function measurements at HERA~\cite{heradata}. The different 
parameters
used in the model are given in Table~\ref{param}. The Sudakov suppression term 
in this
model is given through the hard pomeron characteristics and the gluon radiation 
is thus suppressed thanks to the hybrid  structure of HPM.

%%% ----------------------------------------------------------------------
\subsection{\textbf{The central diffractive cross section }}

All in all, and following the scheme depicted in Fig.\ref{1}, one has the 
following  
matrix element for the central exclusive diffractive production of a massive 
state:
\begin{equation} \label{eq:cs1}
|M|^{2} = (D_{S} G_{S}^{2})^{2} \;
\left([D_{H}G_{H}^{2}]_{1}\right)^{2} \;
\left([D_{H}G_{H}^{2}]_{2}\right)^{2} \;
|M_{\hat{\sigma}}|^{2} \;
%F(|t_{1}|)F(|t_{2}|)]^{2} \;
%\left( \frac{s}{s_{2}} \right)^{2\alpha_{1}-2} \;
%\left( \frac{s}{s_{1}} \right)^{2\alpha_{2}-2}
\end{equation}
The notation $[D_{H}G_{H}^{2}]_{i}$, i=1,2, is used to distinguish the hard 
colored exchanges 
from each vertex, see Fig.\ref{1}.
$M_{\hat{\sigma}}$ is the hard process matrix element for the considered 
produced massive state. 

In parallel with  the approach of Ref.~\cite{bialas1},
the cross section is written as:
% Integrating over the final protons momenta
\begin{eqnarray} \label{eq:cs2}
\sigma & = & 81 \times \frac{2s}{(2\pi)^{5}} 
 \times \left[G_{S}^{2}\ D_{S}^{(0)} \right]^2 \;
\int d^{4} p_{1} d^4 p_{2}~ \delta(p_{1}^{2}) \delta(p_{2}^{2}) \;
\delta((p_{a}\!+\!p_{b}\!-\!p_{1}\!-\!p_{2})^{2} - M^{2}) \times \\ \nonumber 
&\times&\left (\frac{s}{s_1} \right )^{2 \alp (t1) -2}
\left (\frac{s}{s_2} \right )^{2 \alp (t2) -2}
e^{2bt_1} e^{2bt2} \left[G_{H}^{2}\ D_{H}^{(0)} \right]_{1}^2
 \left[G_{H}^{2}\ D_{H}^{(0)} \right]_{2}^2
~|M_{\hat{\sigma}}|^{2} \ .
\end{eqnarray}

Using relation \cite{bialas2}
\begin{equation}
\int d^{4}p_{i} \delta(p_{i}^{2}) = -\frac{1}{2} \int d\xi_{i}\ d^{2}\vec{v_{i}} 
\ 
;\quad 
\frac{s}{s_{i}} = \frac{1}{\xi_{i}}
\la{variables}\end{equation}
where $\vec{v_{i}}$ is the transverse momentum of the final protons and
changing the variable $v_{i}$ to $|t_{i}|$ using $|\vec{v_{i}}|^{2} = 
(1-\xi_{1})|t_{i}|,$ one finally finds:
\begin{eqnarray} \label{eq:cs3}
\sigma & = &
\frac{81}{2(2\pi)^3} \times
\left[G_{S}^{2}\ D_{S}^{(0)} \right]^2 \times
\prod_{i=1,2} \left(
\int \int d\xi_{i} d|t_{i}| 
\frac{1-\xi_{i}}{{\xi_{i}}^{2\epsilon}}
\exp(-(2b+2\alp'\log(\frac{1}{\xi_{i}}) |t_{i}| ) )
\left[G_{H}^{2}\ D_{H}^{(0)} \right]_{i}^2
|M_{\hat{\sigma}}|
\right)~.
\end{eqnarray}

% \eject
%%% ----------------------------------------------------------------------
\section{\textbf{Comparison with Dijet CDF data}}

\subsection{Model implementation in FPMC}
The model has been fully implemented in FPMC~\cite{fpmc}, using the parameters
defined in the previous sections. The different parameters in the hard
part of the model
come mainly from a fit to HERA data (structure function
$F_2$, charm and vector meson data) inspired by saturation models.
By default, we take the parameters from a fit to the diffractive
structure function $F_2$ measured by the H1 and ZEUS collaborations
at HERA~\cite{soyez}. The systematics uncertainties on the fit parameters
define the systematic uncertainties of our model. In addition, it is
possible to include heavy quarks in the model~\cite{soyez}, and compare
it to the vector meson production cross section~\cite{marquet}, which leads
to different parameters of the model (see Table I). The difference of the 
results
with and without including charm effects is also a kind of systematic 
uncertainty in
the model and will be discussed further in the paper. 
In addition, the parameters related to the soft exchange come from
the Donnachie-Landshoff model. All parameters are given for reference 
in Table~\ref{param}. The only parameter in the model is the free
normalisation which we will obtain from a fit to the CDF exclusive diffractive
measurements. Implicitely, the normalisation will thus include the surival 
probability.
Note that the ratio of the survival probabilities between the Tevatron (0.1) and 
the LHC
(0.03) is taken into account when we predict later on the cross sections at the 
LHC.

The implementation in FPMC~\cite{fpmc} allows to interface the hybrid model with 
a jet
algorithm after hadronisation performed in HERWIG~\cite{herwig}. The standard
jet algorithm~\cite{cdfalgo} used by the CDF collaboration has been implemented
so that we are able to compare directly our model with the CDF measurements of
exclusive events.

\begin{table} 
\begin{center}
\begin{tabular}{|c||c||c|c|} \hline
Parameter& Central value & Uncertainties & Charm included \\
\hline\hline
{\bf Hard parameters} & &  & \\
\hline\hline
$N_0$ & 0.7  &  - & 0.7 \\ 
$Q_0$ & 0.254 GeV  & 0.243-0.263 & 0.298 \\ 
$R_p$ & 3.277 GeV$^{-1}$  &  3.233-3.321 & 3.344\\ 
$\gamma_C$ & 0.6194  & 0.6103-0.6285 & 0.7376 \\
$\kappa$ & 9.9  & - & 9.9\\ 
$\lambda$ & 0.2545  &  0.2494-0.2596 & 0.2197\\
$B$ & 2   & - & 2\\ 
$\mu_H$ & 0.5  & - & 0.5 \\ 
\hline
{\bf Soft parameters} &  &  & \\
\hline\hline
$\alpha_P(0)$ & 1.08   & - & 1.08\\ 
$\alpha'$ & 0.06   & - & 0.06\\ 
$\beta$ & 4   & - & 4\\ 
$b$ & 4   & 3-5 & 4\\ 
\hline
\end{tabular}
\end{center}
\caption{{\it List of parameters used in the HPM.} The second column give the
default values used in the model, the third one the range of values used 
for systematics coming from the fit uncertainties
to $F_2$ data, and the fourth one the values of parameters when heavy
quarks are also considered in the model (see text).}
\label{param}
\end{table}

\subsection{Comparison with CDF data}

To test the accuracy of the model it is useful to compare with the CDF
measurements of exclusive events in the dijet channel at the
Tevatron~\cite{cdf,Kepka:2007nr}.
CDF used the dijet mass fraction to quantify the amount of exclusive events.
The dijet mass fraction, namely the ratio of the dijet mass to the total mass in
dijet events, is expected to peak around $1$ for exclusive events since two jets
and nothing else are produced in the final state while inclusive events show
lower values of the dijet mass fraction. The comparison between the CDF 
measurement and what is expected
from inclusive diffraction based from quark and gluon densities measured at HERA
(including the survival probability) leads to an estimate
of the exclusive event cross section. The result is given in Fig.~\ref{cdf1}.
Data points show the exclusive cross section for jets with a transverse momentum
greater than a threshold value given in abscissa. To compare with the
expectation from HPM, the FPMC Monte Carlo was interfaced with
the jet cone algorithm used by the CDF collaboration at hadron level. 
Since the normalisation is not determined by the model, we choose to fix it 
using
the CDF measurement. The global normalisation is obtained by fitting our
predictions to the CDF measurement given in Fig.~\ref{cdf1}.
The normalisation  is found to be: $3.85 \times 10^{-4} \pm 1.89 \times 10^{-4}$ 
with $\chi^2=0.67$ for 5 data points and
the uncertainty comes from the uncertainty on the CDF measurement.
In Fig.~\ref{cdf1}, we give the prediction of HPM in full line, and
the dashed line shows the uncertainty on normalisation ($\pm 1 \sigma$) coming 
from
the fit to the CDF data. We note that the shape of the HPM prediction describes
nicely the CDF data while the normalisation comes directly from the CDF data as
we mentioned previously. 

\begin{figure}
\begin{center}
\epsfig{file=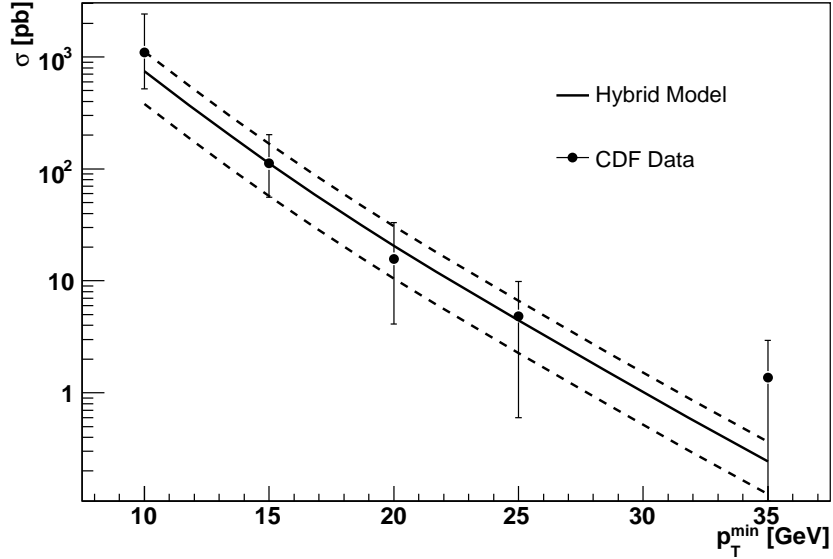, height=0.35\textheight}
\end{center}
\caption{{\it Jet $E_{T_{min}}$ distribution for exclusive events measured by 
the CDF
collaboration compared with the hybrid model.} The shape of the distribution
is well reproduced by the model and the normalisation is fitted to the CDF
measurement ($3.85 ~10^{-4} \pm 1.89 ~10^{-4}$).}
\la{cdf1}
\end{figure}

In Fig.~\ref{cdf2}, we compare the predictions from the hybrid model to the
dijet mass measurements in diffractive exclusive events from the CDF
collaboration. As explained in the CDF paper~\cite{cdf}, this is an indirect
measurement which is MC dependent due to the method used by the CDF 
collaboration
to extract the dijet mass cross section. We follow the same method used by the 
CDF
collaboration to compute the dijet mass cross section. Namely, we convert the measured
exclusive dijet cross section from CDF presented in Fig.~\ref{cdf1}  to a cross 
section versus dijet mass using the HPM. After each $E_{T_{min}}$ cut
(10, 15, 20, 25, and 35 GeV), we normalise the HPM
cross section to the CDF measurement. We have thus a ``calibration" factor
in each $E_{T_{min}}$ interval. The $M_{JJ}$ distribution coming from the hybrid
model is then reweighted after applying the $E_{T_{min}}$ cut using the same
calibration factors. Removing the cuts on $E_{T_ {min}}$ allows to obtain the
``CDF points" given in Fig.~\ref{cdf2}. We followed basically the same procedure
as in Ref~\cite{cdf}, but using the reweighted HPM instead of KMR. 
It is worth noticing that it is not strictly speaking a measurement by the
CDF collaboration since it is model dependent. Nevertheless, we can now compare
the ``CDF measurement" to the expectation of the hybrid model and the result is
shown in Fig.~\ref{cdf2}.
The dashed line indicates the uncertainties on the model
related to the normalisation. The model leads to a good description of CDF data
over the full dijet mass range.

\begin{figure}
\begin{center}
\epsfig{file=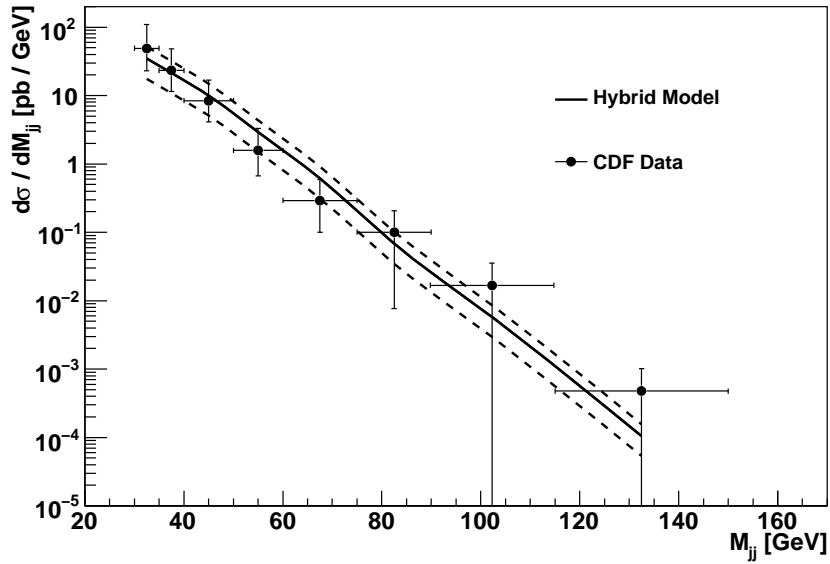, height=0.35\textheight}
\end{center}
\caption{{\it Dijet mass distribution for exclusive events using the CDF method
compared to the hybrid model.} The normalisation comes from the fit to the
CDF $E_{T_{min}}$ distribution (see Fig.~\ref{cdf1}) and the shape is well
described by the hybrid model.}
\la{cdf2}
\end{figure}

\subsection{Uncertainties on the model predictions}
In this section, we discuss the uncertainties related to the chosen values of
parameters given in Table~\ref{param}. The first uncertainties come from the
uncertainties on the parameter used to describe the hard interaction. As
we mentioned already, the values of the parameters are taken from a fit to $F_2$ 
data
coming from the HERA experiments~\cite{soyez}. The values
of the parameters found in Ref.~\cite{soyez} where obtained with a given
uncertainty coming from the fit procedure and it is worth checking the effect on
the HPM predictions. There was also another kind of fits performed in
Ref.~\cite{soyez} where heavy quarks were considered 
and we also compare our predictions including or not the heavy quarks.
The values of the parameters are given in Table~\ref{param} for references.
It is worth noticing that we use the same values of parameters coming from a fit 
to HERA
data to extrapolate at LHC energies, especially when we predict the exclusive 
Higgs boson
cross section. It will be thus important to test the values of the parameters 
using directly
LHC data when they will be available, and to study whether this assumption is 
valid.
The effect of changing the hard parameters are given in Fig.~\ref{models1} for 
the
jet $E_{T_{min}}$ and the $M_{JJ}$ distributions. The differences are found to 
be less than 20
\%. 

Another systematic study we performed was to change the $b$ slope of the
soft cross section responsible for the soft interaction. The uncertainty on the
$b$ slope coming from soft data is quite small but we wanted to study the 
dependence
of our model on this parameter.
Modifying the $b$
parameter from 2 to 4 leads to the cross sections given in Fig.~\ref{models3} 
for the
jet $E_{T_{min}}$ and the $M_{JJ}$ distributions. The difference is found to be 
less than 20\% everywhere.
It is worth noticing that the leading uncertainty in the predictions for HPM 
comes from the statistical uncertainties of the $E_{T_{min}}$ cross
section measurement by the CDF collaboration which is of the order of 50\%.

The effect of taking the parameters of the fit of Ref.~\cite{soyez} where heavy
quarks are considered are given in Fig.~\ref{models5}.
We recomputed the normalisation by fitting the $E_{T_{min}}$ distribution to the
CDF data and the normalisation for the light quark only model is 6.80 $\times$ 10$^{-3}$
$\pm$  3.46 $\times$ 10$^{-3}$ with a $\chi^2$ of 0.83 for 5 data points.
We notice that the mass dependence is stronger when heavy quarks are
considered, which means that the cross section at high mass is 
slightly smaller, and that the fit to the CDF data on $E_{T_{min}}$ is slightly 
worse.

\begin{figure}
\begin{center}
\epsfig{file=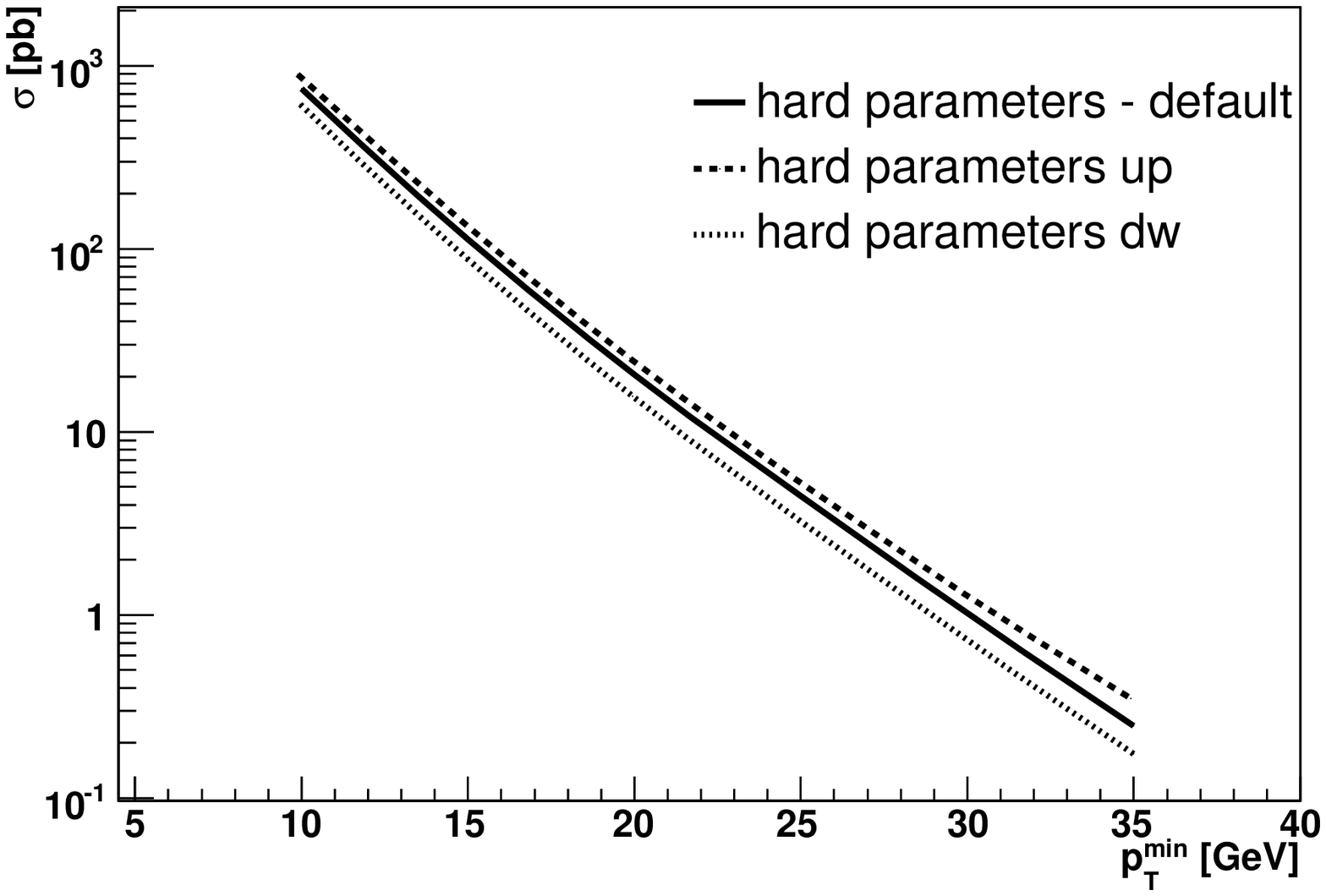, height=0.25\textheight}
\epsfig{file=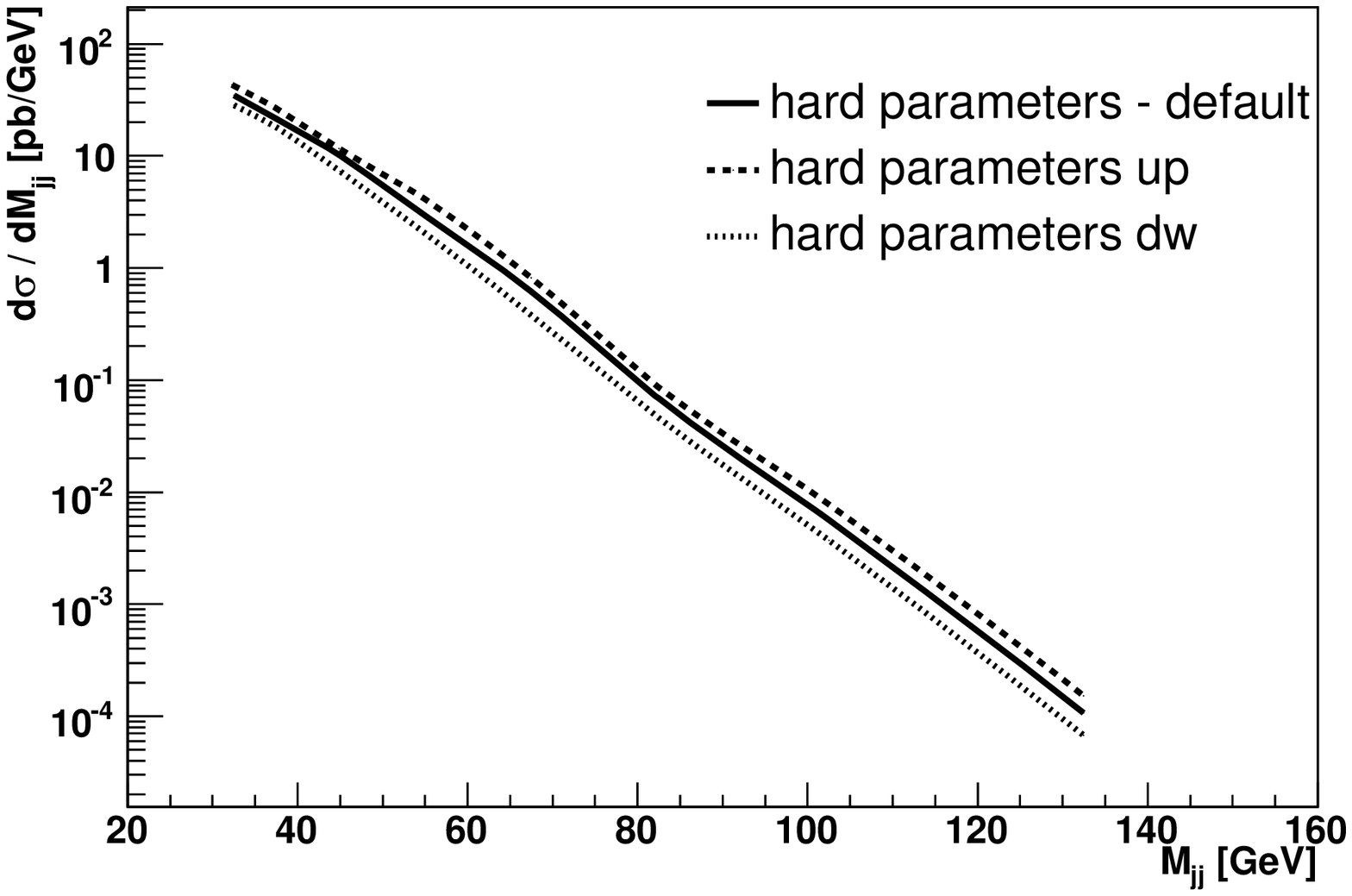, height=0.25\textheight}
\end{center}
\caption{{\it Effect on modifying the hybrid model hard parameters on the 
$E_{T_{min}}$ and $M_{JJ}$ cross section distributions.}}
\la{models1}
\end{figure}

\begin{figure}
\begin{center}
\epsfig{file=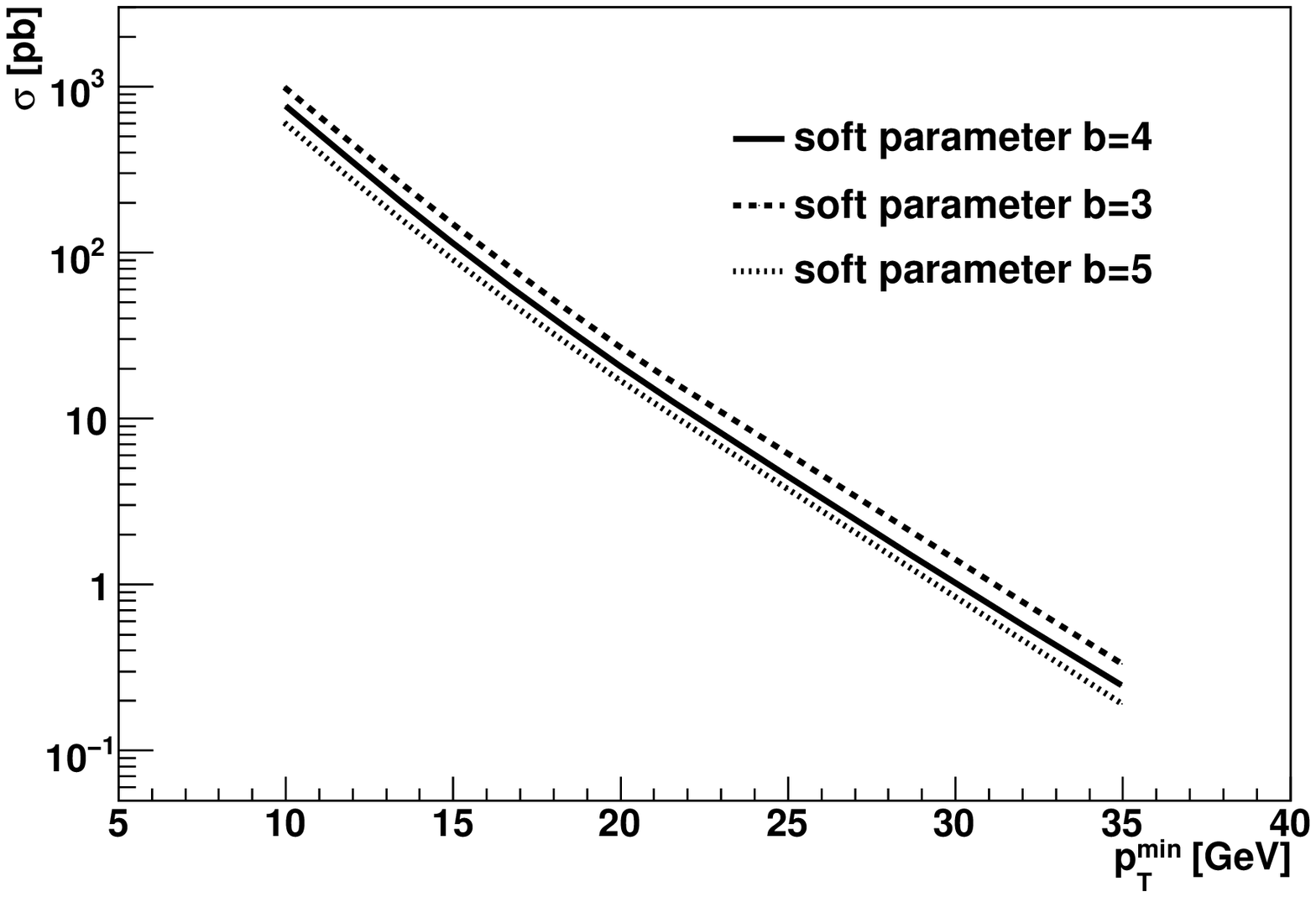, height=0.25\textheight}
\epsfig{file=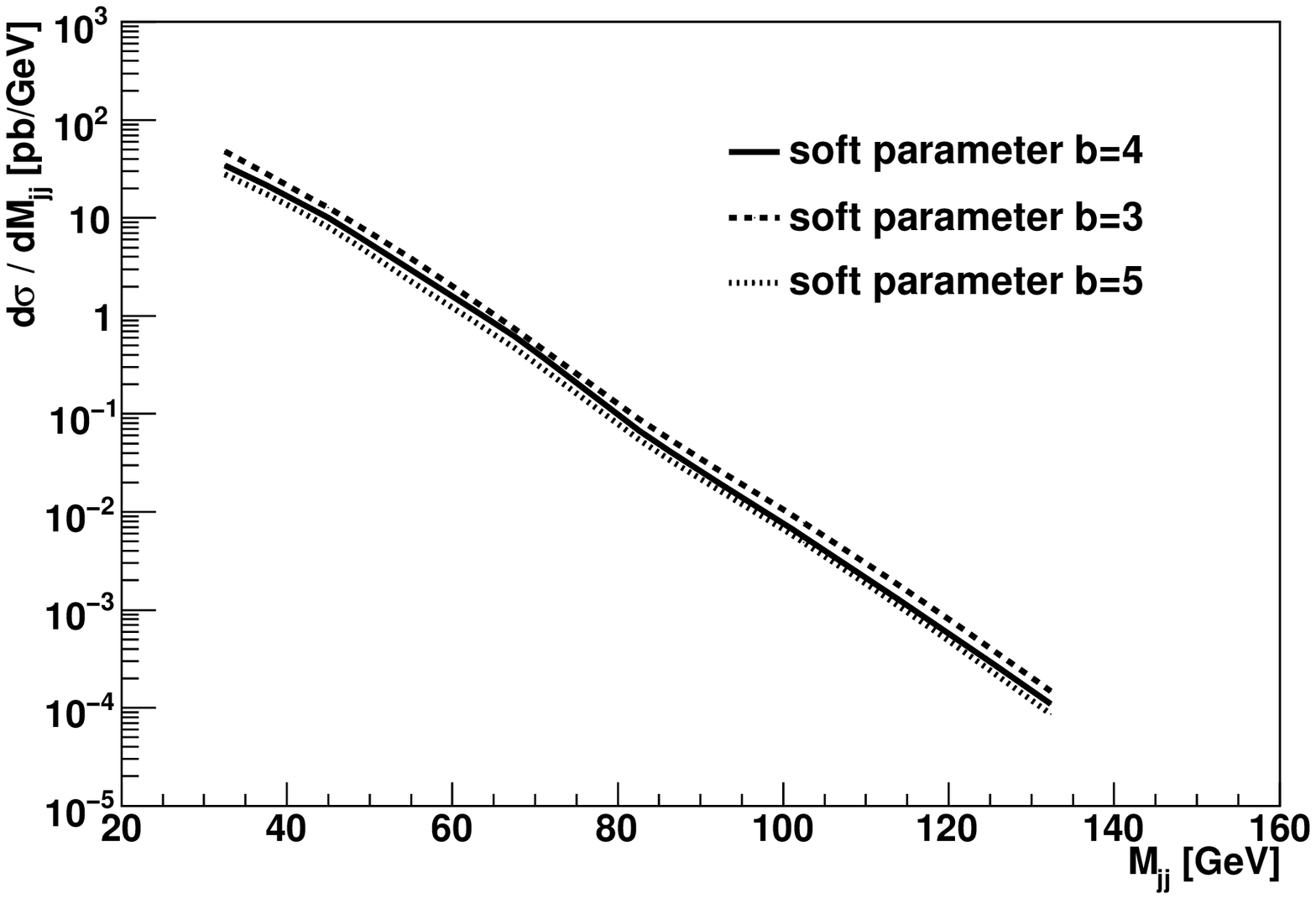, height=0.25\textheight}
\end{center}
\caption{{\it Effect on modifying the hybrid model soft $b$ parameters on the 
$E_{T_{min}}$ and $M_{JJ}$ cross section distributions.}}
\la{models3}
\end{figure}

\begin{figure}
\begin{center}
\epsfig{file=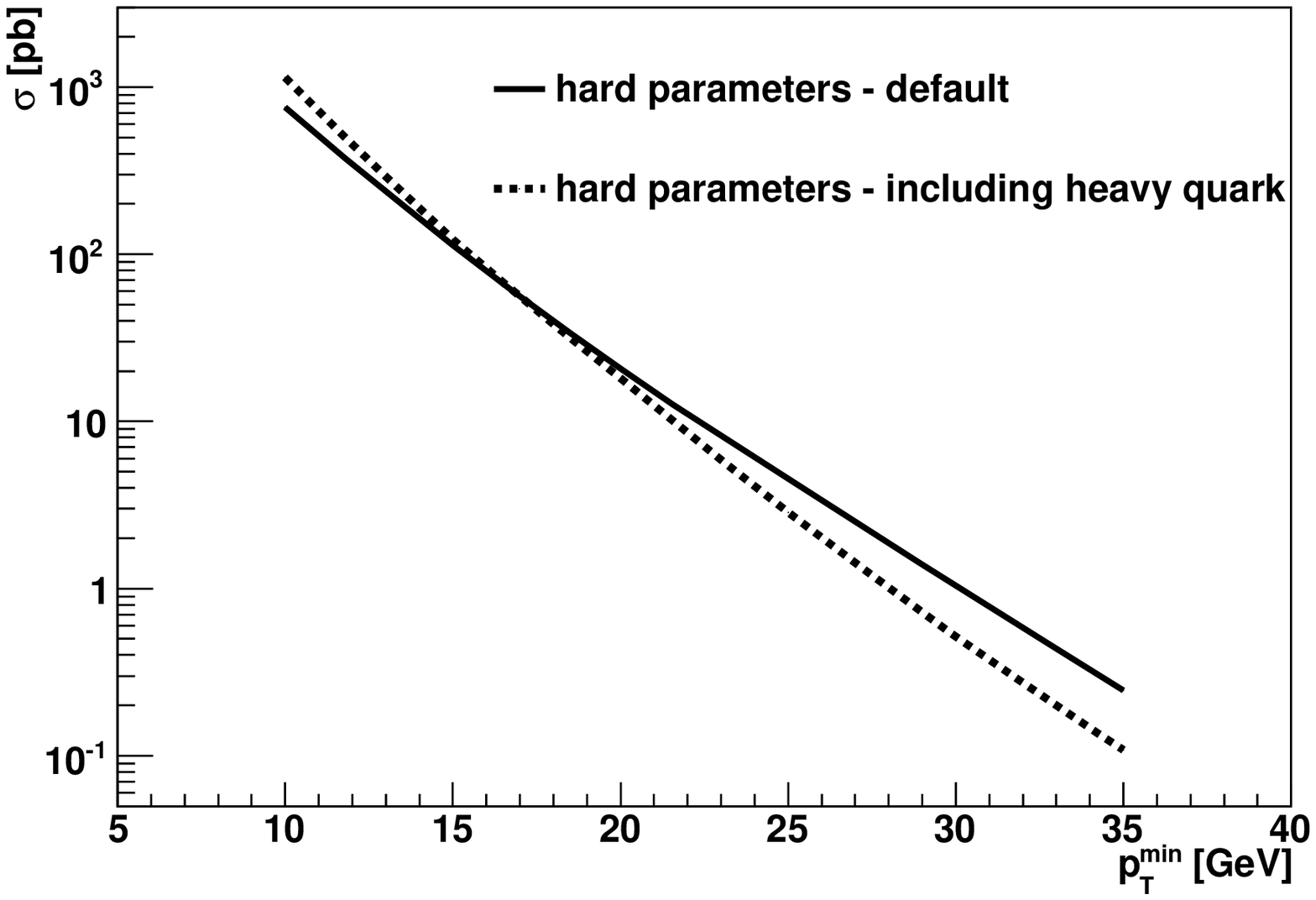, height=0.25\textheight}
\epsfig{file=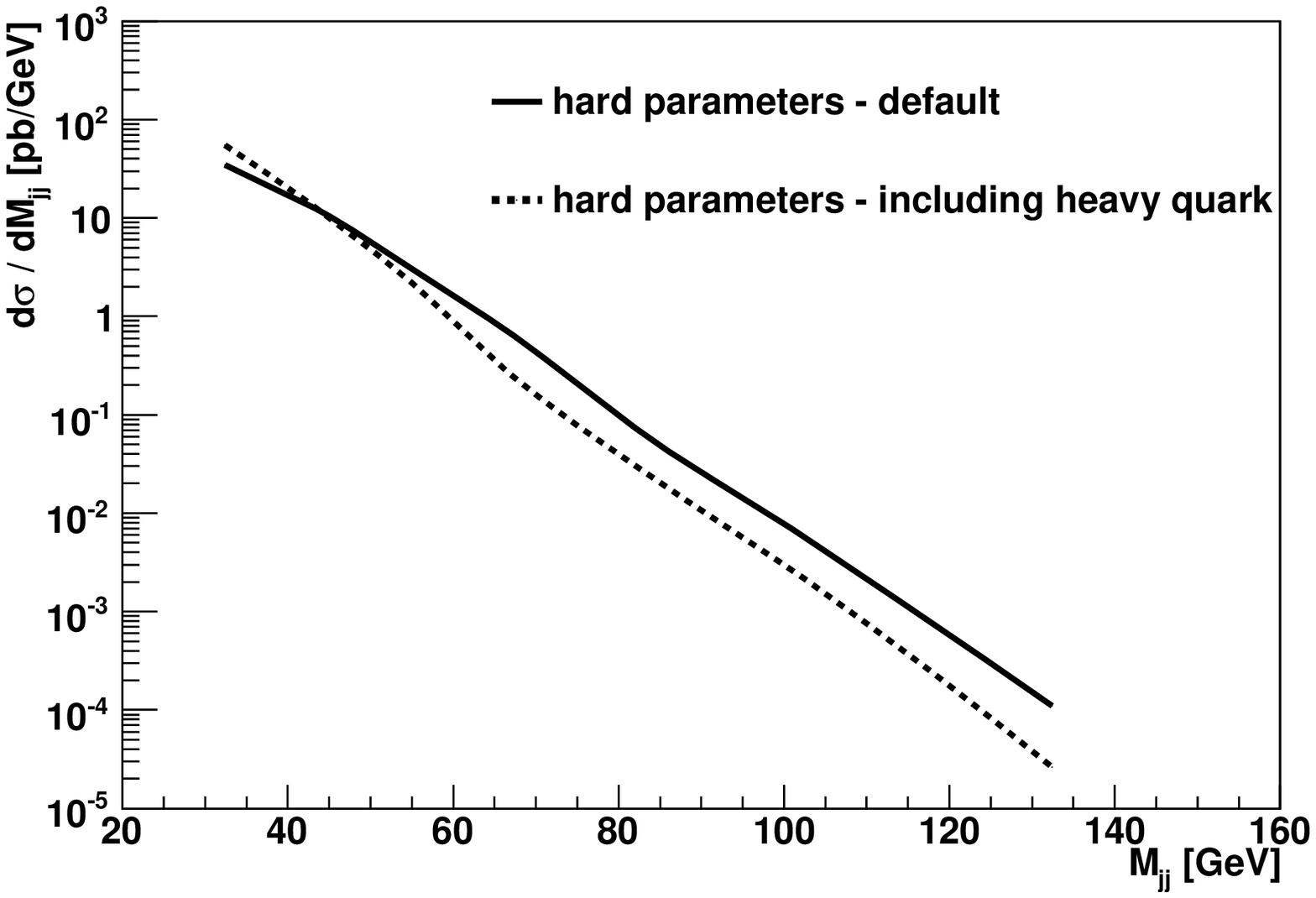, height=0.25\textheight}
\end{center}
\caption{{\it Comparison of the jet $E_{T_{min}}$ and $M_{JJ}$ distributions for 
exclusive events 
with the hybrid model including or not heavy
quark effects.} The normalisation comes from a fit to the CDF exclusive 
$P_{T_{min}}$ cross
section measurements. We note that including heavy quarks leads to a stronger 
$P_{T_{min}}$ 
dependence.}
\la{models5}
\end{figure}

\subsection{\textbf{Predictions for the LHC}}
In Fig.~\ref{lhc1}, we show the exclusive Higgs boson cross section using the
HPM. The cross section varies from 1.1 $\pm$ 0.5 fb at 120 GeV to 0.32
$\pm$ 0.15 fb at 160 GeV. Including heavy quark effects reduces this
cross section by about 60\%.
The values are found to be slightly lower than with the KMR model but
compatible within uncertainties, and we should also notice that these
predictions are at LO and it is known that NLL corrections increase the cross
section of typically about 20\%.

In Fig.~\ref{lhc2}, we also compare the $\xi$
distributions for jet production in exclusive events for 
the HPM and KMR models for jets with $p_T>$50
at the LHC. The $\xi$-slope is found 
to be smoother at the LHC for the KMR model than for the HPM. 
LHC data should thus
allow to distinguish between both models or to tune better the parameters of the
HPM given in the previous section.

\begin{figure}
\begin{center}
\epsfig{file=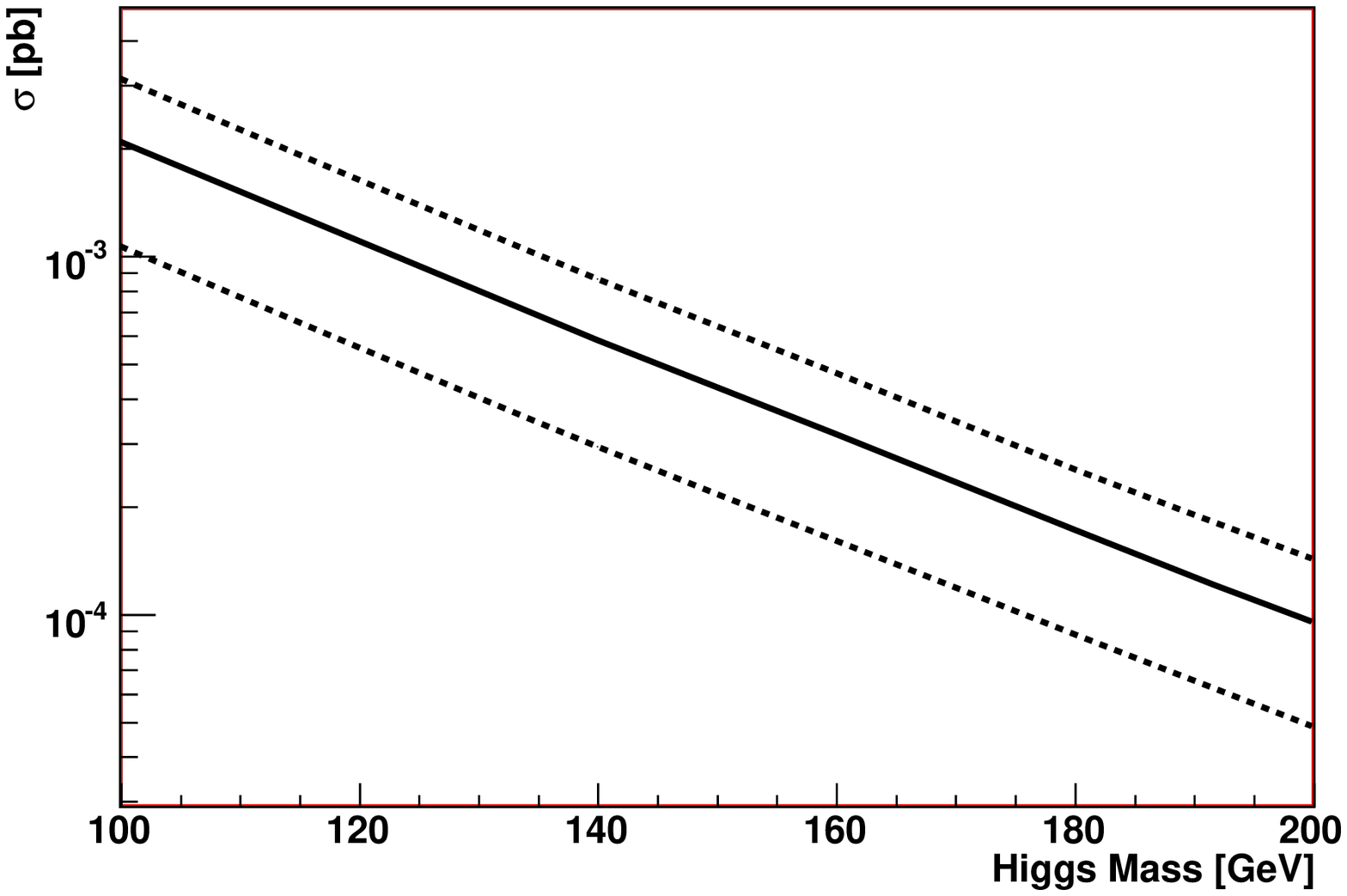, height=0.35\textheight}
\end{center}
\caption{{\it Prediction on the diffractive exclusive Higgs cross section at the 
LHC
using the HPM.}}
\la{lhc1}
\end{figure}

\begin{figure}
\begin{center}
\epsfig{file=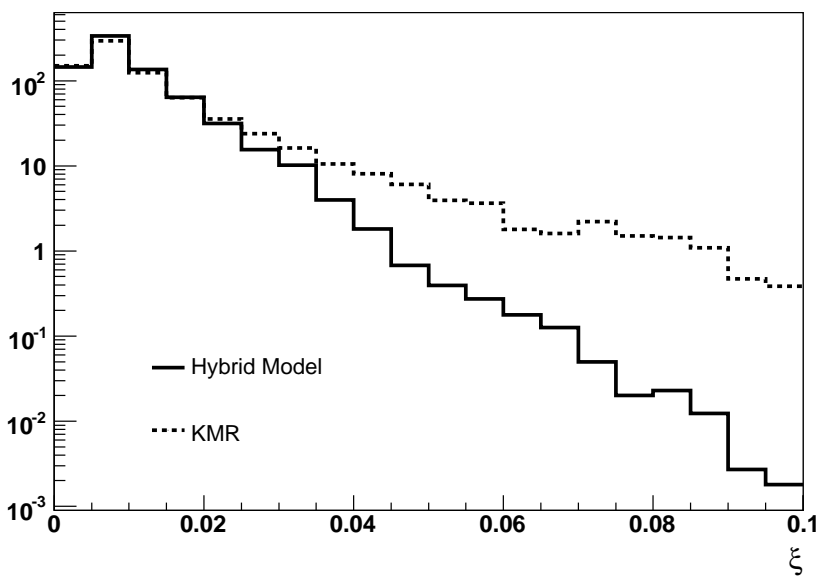, height=0.35\textheight}
\end{center}
\caption{{\it $\xi$ distribution for exclusive events for jets with $p_T>$ 50 
GeV at the LHC 
for the KMR and HPM models.} The $\xi$-dependence is smoother for KMR 
than for the 
HPM model. LHC data will help to disctinguish and tune both models.}
\la{lhc2}
\end{figure}

\section{Conclusion}
%\section*{Appendix}
In this paper, we propose a new model to describe exclusive event production at 
hadronic
colliders. It is based on Double Pomeron Exchange. We call it  ``Hybrid
Pomeron Model" (HPM) since one of the color exhanges is considered to be hard, 
taking away most of the transverse momentum 
available while the colorless
aspect of the overall crossed channel is ensured via a soft additional color 
exchange. The parameters 
of the model
come from a fit to HERA $F_2$ data using a BFKL-based model for the hard part 
(eventually including saturation corrections), 
while the parameters
of the soft part come from the usual soft cross section models. The model was 
successfully
implemented in a generator (FPMC) to be able to compare directly with the CDF 
measurements performed
at particle level. 

Our predictions are found to be in very good agreement with the measurements of 
the 
exclusive cross section as a function of the minimum jet transverse momentum
or the dijet mass from the CDF collaboration. The HPM predicts a Higgs boson 
production cross section of about 1.1 fb  
at the LHC for a Higgs boson mass of 120 GeV. This is in the same range and 
compatible with the KMR 
determination.
The $\xi$ distribution for exclusive events is softer for KMR than 
for the HPM model and it
will be worth measuring it at the LHC and the Tevatron to distinguish and 
further tune both models.
As we mentionned, the parameters used in the HPM come from an extrapolation from 
a fit to
HERA data and it will be good to cross check the values of the parameters using 
direct data from the
LHC.

\acknowledgments

We thank Cyrille Marquet for useful remarks. One of us (M.S.R.) acknowledges 
support from CNPq (Brazil).

\eject

%%% ----------------------------------------------------------------------

\end{document}